\documentclass{article}

\usepackage[main, final]{neurips_2026}

\usepackage[utf8]{inputenc} 
\usepackage[T1]{fontenc}    
\usepackage{hyperref}       
\usepackage{url}            
\usepackage{booktabs}       
\usepackage{amsfonts}       
\usepackage{nicefrac}       
\usepackage{microtype}      
\usepackage{xcolor}         
\usepackage{amsmath}
\usepackage{amssymb}
\usepackage{graphicx}

\title{Direct Conditioning of Audio Diffusion Transformers on fMRI Reveals Cortical Contributions to Sound Reconstruction}

\author{%
  \parbox[t]{0.44\textwidth}{\centering\normalfont
    \textbf{Matteo Ciferri}\\
    Department of Biomedicine\\
    and Prevention\\
    University of Rome Tor Vergata, Italy\\
    {\small\texttt{matteo.ciferri@students.uniroma2.eu}}
  }
  \And
  \parbox[t]{0.44\textwidth}{\centering\normalfont
    \textbf{Tonio Weidler}\\
    Maastricht Centre for\\
    Systems Biology (MaCSBio)\\
    Faculty of Science and Engineering\\
    Maastricht University, The Netherlands
  }
  \AND
  \parbox[t]{0.44\textwidth}{\centering\normalfont
    \textbf{Matteo Ferrante}\\
    Tether Evo, El Salvador \\
    Department of Biomedicine\\
    and Prevention\\
    University of Rome Tor Vergata, Italy
  }
  \And
  \parbox[t]{0.44\textwidth}{\centering\normalfont
    \textbf{Nicola Toschi}\thanks{These authors contributed
    equally as senior authors.}\\
    Department of Biomedicine\\
    and Prevention\\
    University of Rome Tor Vergata, Italy\\
    Athinoula A. Martinos Center\\
    for Biomedical Imaging\\
    Massachusetts General Hospital\\
    and Harvard Medical School, USA
  }
  \AND
  \parbox[t]{0.65\textwidth}{\centering\normalfont
    \textbf{Elia Formisano}\footnotemark[1]\\
    Maastricht Brain Imaging Center\\
    Department of Cognitive Neuroscience\\
    Faculty of Psychology and Neuroscience\\
    Maastricht University, The Netherlands
  }
}

\begin{document}

\maketitle

\begin{abstract}
  Reconstructing audio from brain activity remains challenging. \textcolor{black}{Existing brain-to-audio pipelines typically map neural responses to predefined stimulus representations before synthesis. While recent work in the visual domain has explored direct conditioning on brain activity, this approach has not yet been extended to audio diffusion models. We introduce MAAS (Multi-Area Auditory Synthesis), an end-to-end framework that conditions a pretrained audio diffusion transformer on ROI-wise fMRI responses without first supervising a decoder to predict an external stimulus representation. MAAS uses region-specific projections and learned brain tokens to condition Stable Audio through a dedicated cross-attention pathway.} On a 7T fMRI dataset, we demonstrate consistent improvements over baselines, including two-stage fMRI-to-latent conditioning, across both acoustic and semantic evaluation metrics. In addition, we introduce a fidelity metric grounded in a computational model of auditory neural processing, which allows us to quantify the contribution of individual auditory cortical regions to reconstructed audio representations.
\end{abstract}

\section{Introduction}

Reconstructing audio from functional Magnetic Resonance Imaging (fMRI) measurements of brain activity offers a unique window into the representational structure of auditory cortex, yet remains technically challenging. While recent advances in the visual domain have demonstrated that photorealistic images can be synthesized from fMRI \citep{ferrante2024decoding,Ozcelik2023,luo2023brain,chen2023cinematic}, progress in the auditory domain has been comparatively limited. This gap reflects several compounding factors: the hemodynamic response introduces temporal blurring on the order of seconds \citep{kay2013glmdenoise}, auditory datasets with sufficient diversity remain scarce, and sound reconstruction imposes strict spectrotemporal constraints. 

Audio generative models impose a restrictive conditioning interface, typically accepting semantic embeddings (e.g., text or audio-language representations). \textcolor{black}{Recent work tried to extend text-conditioned audio generation with additional, non-text conditioning signals. AP-Adapter \citep{tsai2024audio} augments AudioLDM2 \citep{liu2024audioldm} with a parallel cross-attention pathway, allowing the model to use acoustic information alongside the text prompt for sound editing. MuseControlLite \citep{musecontrollite} applies decoupled cross-attention to Stable Audio \citep{evans2025stable} to incorporate time-varying musical attributes, such as melody, rhythm, and dynamics, as well as reference-audio signals, enabling temporally controlled generation and audio inpainting or outpainting.}

Existing brain-to-audio pipelines address the conditioning mismatch through a two-stage strategy, mapping neural activity onto intermediate deep neural network (DNN) embeddings, such as CLAP \citep{elizalde2023clap} or YamNet \citep{hershey2017cnn}, which are then used to guide a generative model. While effective, this approach introduces a key limitation: it constitutes an information bottleneck \citep{shirakawa2025spurious, feng2025towards}, and reliance on intermediate embeddings may discard fine-grained spectrotemporal structure. Recent work in decoding of visual stimuli from brain activity has begun to explore an alternative paradigm based on direct conditioning, where generative models are conditioned on brain representations without intermediate embedding prediction \citep{feng2025towards}. Related ideas have also been explored in language domain, where neural signals are directly integrated into autoregressive generation \citep{ye2025generative}. However, these approaches are tailored to specific architectures and do not transfer directly to current audio diffusion models. 

\textcolor{black}{Here, we introduce MAAS (Multi-Area Auditory Synthesis), an end-to-end framework for conditioning a pretrained audio diffusion transformer on ROI-wise fMRI responses.} Central to our approach is a transformer-compatible adapter that injects neural representations into the cross-attention layers of a diffusion transformer (DiT), while remaining compatible with pretrained models such as Stable Audio \citep{evans2025stable}.  Importantly, MAAS preserves anatomical organization at the region-of-interest (ROI) level by conditioning on ROI-specific representations. This enables an interpretable mapping between brain regions and generated audio, allowing the contribution of distinct cortical areas to be explicitly probed. This is important because different regions of auditory cortex are thought to encode complementary aspects of sound, and separating their contributions helps reveal how complex auditory representations are organized in the brain. 

We instantiate this framework for brain-to-audio reconstruction from fMRI. Early studies demonstrated that spectrotemporal features of natural sounds could be predicted from fMRI patterns using multivariate regression \citep{santoro2017reconstructing}, though resulting reconstructions lacked perceptual fidelity. More recent approaches have improved realism by decoding high-level DNN features and mapping them to a codebook-based audio generator \citep{park2025natural}. \cite{denk2025text} and \cite{ciferri2025reconstructing} have pursued decoding of musical stimuli from neural signals, both routing through latent representations before synthesis. In these pipelines, the intermediate representation also serves as a denoising stage, leveraging regularization to suppress measurement noise before synthesis \citep{ferrante2023eyes}.

In contrast, our setting leverages the signal-to-noise ratio of fMRI responses acquired at ultra-high magnetic field strength (7 Tesla) \citep{santoro2017reconstructing, giordano2023intermediate}, which exceeds that of conventional field strengths. This enables robust estimation of voxel-wise responses to individual sounds, substantially reducing measurement noise. 

MAAS reconstructs natural sounds from fMRI with improved acoustic and semantic fidelity. The structured ROI-based conditioning enables analysis of the representational contribution of individual cortical regions via an ablation-based fidelity metric computed in a biologically grounded auditory representation space. We further characterize how brain-derived signals are guided through the generative process via attention routing.

\textcolor{black}{Our contributions are: (i) an end-to-end framework for conditioning a pretrained audio diffusion transformer on ROI-wise fMRI responses, without an intermediate decoder trained to predict a predefined stimulus representation; (ii) an evaluation on a 7T dataset, including controlled two-stage and caption-based baselines, and a neural-information controls; (iii) an analysis of ROI-level perceptual contributions using a biologically grounded auditory representation.}

\section{Related Work}
\label{rel_works}
\paragraph{Brain decoding with generative models.}
Recent work has shown that perceptual stimuli can be reconstructed from fMRI by conditioning generative models on features decoded from brain activity. In vision, latent diffusion models combined with DNN feature decoding (e.g., CLIP\citep{radford2021learning}, DINO \citep{oquab2023dinov2}) have enabled high-quality image reconstruction \citep{ferrante2024decoding,takagi2023high,scotti2024mindeye2}. In the auditory domain, early work decoded spectrotemporal features directly \citep{santoro2017reconstructing}, while more recent approaches using audio-AI models have similarly relied on intermediate representations to drive generative models \citep{park2025natural,denk2025text,ciferri2025reconstructing}. In contrast, our approach removes this intermediate stage by directly conditioning the generative model on neural signals.

\paragraph{Direct conditioning on neural representations.}
A small number of recent works have explored conditioning generative models directly on brain activity. NeuroAdapter \citep{feng2025towards} conditions a visual latent diffusion model, based on a UNet architecture \citep{rombach2022high}, on parcellated fMRI representations bypassing intermediate feature spaces. Our work extends this paradigm to audio diffusion transformers, which differ in architecture and conditioning interfaces.


\paragraph{Conditioning mechanisms for diffusion models.}
Several methods have been proposed to incorporate auxiliary inputs into diffusion models. IP-Adapter \citep{ye2023ip} introduces a lightweight mechanism for injecting external representations into cross-attention layers via key-value modulation, enabling flexible conditioning beyond text. \textcolor{black}{AP-Adapter and MuseControlLite \citep{musecontrollite} extend text-conditioned audio generation by incorporating reference-audio features or time-varying musical controls via decoupled cross-attention.} We extend the adapter concept to diffusion transformer (DiT)–based audio models, enabling conditioning on continuous signals such as fMRI. 

\paragraph{Interpreting brain contributions.}
While prior decoding work focuses primarily on reconstruction quality, fewer studies address how different brain regions contribute to the generated outputs \citep{santoro2017reconstructing}. By conditioning on ROI–specific signals, our framework enables direct analysis of regional contributions through ablation, providing a more interpretable link between neural activity and generated audio.

\section{Methods}
\label{methods}

\begin{figure*}
    \centering
    \includegraphics[width=.99\linewidth]{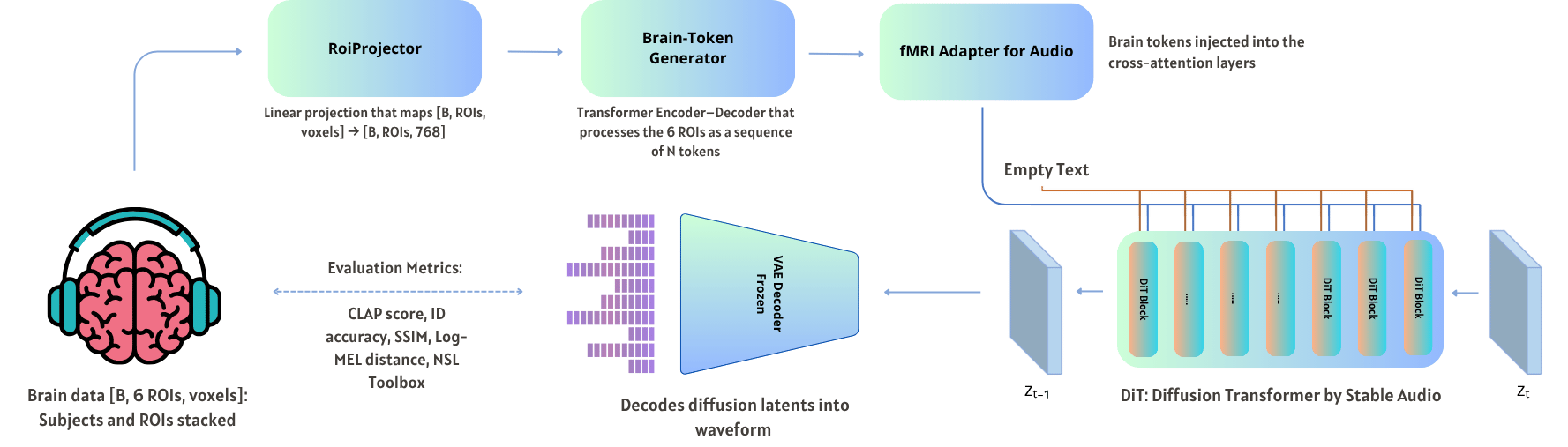}
    \caption{General overview of our approach. fMRI responses from auditory ROIs are projected and transformed into brain tokens, which are injected into the cross-attention layers of a Diffusion Transformer via a dedicated adapter. The model is conditioned solely on brain activity (empty text prompt) and generates audio through a frozen latent diffusion backbone.}
    \label{fig:BrainSounds_overview}
\end{figure*}

We describe the neural data, the voxel selection procedure, the generative backbone, and the architecture of MAAS. Our framework is depicted in Figure \ref{fig:BrainSounds_overview}.

\subsection{Experimental Setup} 
We use the 7T fMRI dataset from \cite{santoro2017reconstructing, giordano2023intermediate}. Five participants listened to 288 natural sounds (1\,s duration each, divided into four non-overlapping sets of 72 sounds each) spanning six semantic categories (human non-speech vocalizations, speech, animal cries, musical instruments, nature scenes, and tool sounds) while fMRI data were acquired using a clustered EPI sequence (TR\,=\,2600\,ms, TA\,=\,1200\,ms, TE\,=\,19\,ms, voxel size\,=\,1.5\,mm isotropic, 7T Siemens). Stimuli were presented in the silent gap between volume acquisitions in a fast event-related design. Each sound was repeated three times across two scanning sessions. 
Single-sound BOLD responses were estimated via a two-step procedure a within four-fold cross-validation across the four stimulus sets (3 sets for training, 1 set for testing). First, using the training data, a voxel-specific hemodynamic response function (HRF) was estimated using finite impulse response (FIR) modeling \citep{kay2013glmdenoise, santoro2014encoding}. Second, sound-evoked responses were estimated by fitting fMRI responses, separately for training and test data, using the estimated HRFs and regularised linear regression (ridge). This procedure yields one response estimate (beta) per sound per voxel. Combined with the high functional contrast-to-noise ratio of 7T fMRI, this procedure provides a robust representation of auditory cortical activity \citep{santoro2017reconstructing}. 

For each participant, responses were extracted from six bilateral (left and right hemispheres combined) anatomical ROIs covering the auditory cortex: Heschl's Gyrus (HG), Planum Temporale (PT), Planum Polare (PP), and posterior, middle, and anterior Superior Temporal Gyrus (pSTG, mSTG, aSTG). Each ROI was manually delineated on individual cortical surfaces in the original study.
The final input to MAAS is a matrix $\mathbf{X} \in \mathbb{R}^{R \times V}$ per sample, where $R = 6$ is the number of ROIs and $V = 1024$ is the number of selected voxels per ROI (see below). 


Since the number of voxels per ROI differs across participants, direct pooling across subjects is not possible. We address this by selecting the $V = 1024$ most responsive voxels per ROI per participant, quantified as the prediction accuracy (Pearson $r$) of a ridge regression encoding model trained to predict voxel responses from YamNet embeddings \citep{hershey2017cnn}, fitted exclusively on the training split (in cross-validation) of each fold to avoid circularity. This yields fixed-dimensional ROI representations that can be pooled across participants and fed jointly to MAAS, with the first module (Projector) learning shared weights across subjects. Note that YamNet serves solely as a voxel selection criterion, not as a conditioning signal for generation.

\subsection{Generative Backbone: Stable Audio}

We build on Stable Audio \citep{evans2025stable}, a latent diffusion model designed for text-conditioned audio synthesis. The model operates in a compressed latent space obtained via a variational autoencoder (VAE), which maps waveforms $\mathbf{w} \in \mathbb{R}^{1 \times T_w}$ to latent representations
$
    \mathbf{z} = \mathcal{E}(\mathbf{w}) \in \mathbb{R}^{C \times T_z},
$
where $C$ is the number of latent channels and $T_z$ is the temporal length of the latent sequence. A Diffusion Transformer (DiT) performs iterative denoising in this latent space, conditioned on textual inputs encoded by the flan T5 model \citep{chung2024scaling}. The resulting conditioning sequence $\mathbf{c}_\text{text} \in \mathbb{R}^{L \times D}$ (with $L=128$ token length, $D=768$ feature dimension) is injected via cross-attention layers, while duration information is provided through global conditioning embeddings.

A key limitation of Stable Audio, and audio diffusion models more broadly, is that conditioning is restricted to text embeddings. The architecture does not provide a native mechanism for incorporating structured, non-linguistic signals such as neural activity. While auxiliary conditioning methods have been explored in image diffusion models (e.g., NeuroAdapter \citep{feng2025towards}), these approaches are tightly coupled to UNet-based architectures \citep{rombach2022high} and do not transfer directly to transformer-based diffusion backbones. In particular, Stable Audio’s DiT departs from UNet-based designs in several key aspects: it applies rotary positional embeddings (RoPE) \citep{su2024roformer} only to latent self-attention queries, uses grouped-query attention (GQA) \citep{ainslie2023gqa} with fewer key/value than query heads, and exposes conditioning solely as token-level cross-attention rather than spatial feature maps. These choices break the assumptions of UNet-oriented adapters such as IP-Adapter, which rely on spatial attention blocks and matched key-value head structure. Direct reuse is therefore non-trivial: non-text tokens must be separated from text tokens, projected with compatible key/value heads, and routed without disrupting the pretrained denoising interface. MAAS addresses this gap with a dedicated brain pathway for DiT cross-attention.

\subsection{MAAS Architecture}
\label{sec:model}

Our goal is to map fMRI signals directly into the conditioning space of the generative model, without relying on intermediate embeddings. MAAS comprises three components: (i) a ROI Projector that maps voxel-level activity to a shared embedding space, (ii) a Brain Token Generator that transforms region-wise representations into a compact sequence of conditioning tokens, and (iii) a transformer-compatible adapter that injects these tokens into the diffusion backbone. During training, the standard text pathway is neutralized by providing an empty prompt to the T5 encoder. This ensures that all generative information is conveyed through the brain-derived conditioning stream, enabling a direct analysis of how neural signals influence synthesis.

\paragraph{The ROI Projector.}
For each ROI $r \in \{1, \ldots, R\}$ (with $R = 6$ bilateral auditory regions), the corresponding voxel activity vector $\mathbf{x}_r \in \mathbb{R}^{V}$ (with $V = 1024$ selected voxels) is projected to the backbone feature space with dimension $D = 768$ via a region-specific linear map:
$
    \mathbf{h}_r = \mathbf{x}_r \mathbf{W}_r + \mathbf{b}_r \in \mathbb{R}^D,
$
where $\mathbf{W}_r \in \mathbb{R}^{V \times D}$ and
$\mathbf{b}_r \in \mathbb{R}^D$ are learnable parameters. Stacking across regions yields $\mathbf{H} \in \mathbb{R}^{R \times D}$. Unlike approaches that collapse brain activity into a single global representation, this design preserves region-specific structure and allows the model to learn heterogeneous mappings across anatomically distinct auditory areas.

\paragraph{The Brain Token Generator.}
Because the number of ROIs is small, directly conditioning on $\mathbf{H}$ would limit expressive capacity. To address this, we transform the ROI embeddings into a richer set of conditioning tokens using a learned query-based aggregation mechanism.

Specifically, we employ a transformer decoder with learned queries:
\begin{equation}
    \mathbf{C}_\text{brain} = \text{CrossAttn}\!\left(
        \mathbf{Q}_\text{learn},\, \mathbf{H} + \mathbf{E}_\text{roi}
    \right) \in \mathbb{R}^{Q \times D},
\end{equation}
where $\mathbf{Q}_\text{learn} \in \mathbb{R}^{Q \times D}$ are $Q$ learnable query embeddings and $\mathbf{E}_\text{roi} \in \mathbb{R}^{R \times D}$ are learned positional embeddings encoding ROI identity.

This mechanism maps a small set of anatomically grounded signals into a higher-capacity token sequence compatible with the diffusion transformer. Learned queries allow flexible aggregation of information across regions, rather than relying on fixed pooling schemes.

\paragraph{Conditioning the DiT: fMRI Adapter for Audio.}
We integrate brain-derived tokens into the diffusion process through a parallel attention pathway. In Stable Audio, each cross-attention layer computes:
$
    \mathbf{h}' = \mathrm{Attn}(\mathbf{q},\, \mathbf{k}_\text{text},\, \mathbf{v}_\text{text}),
$
where $\mathbf{q} = \mathbf{h}\mathbf{W}^Q$,
$\mathbf{k}_\text{text} = \mathbf{c}_\text{text}\mathbf{W}^K$,
$\mathbf{v}_\text{text} = \mathbf{c}_\text{text}\mathbf{W}^V$.

$\mathbf{q}$ are derived from latent features, and keys/values are obtained from the text conditioning sequence.

In our setting, $\mathbf{c}_\text{text}$ is fixed to the T5 encoding of an empty string. We introduce a parallel "brain" branch with dedicated key and value projections:
$\mathbf{k}_\text{adap} = \mathbf{C}_\text{brain}\,\mathbf{W}^K_\text{adap},$
$\mathbf{v}_\text{adap} = \mathbf{C}_\text{brain}\,\mathbf{W}^V_\text{adap}.$

The output of the attention layer becomes a weighted combination of the two branches:
$
    \mathbf{h}' = \mathrm{Attn}(\mathbf{q},\, \mathbf{k}_\text{text},\, \mathbf{v}_\text{text})
    \;+\; \lambda \cdot \mathrm{Attn}(\mathbf{q},\, \mathbf{k}_\text{adap},\, \mathbf{v}_\text{adap}),
$
where $\lambda$ is a learnable scalar that calibrates the strength of the brain pathway against the pretrained stream, since their optimal relative scale is not known a priori.

To preserve compatibility with the original model interface, we construct an augmented conditioning sequence
\begin{equation}
    \tilde{\mathbf{c}} = \left[\mathbf{c}_\text{empty};\;
    \mathbf{c}_\text{start};\; \mathbf{c}_\text{end};\;
    \lambda \cdot \mathbf{C}_\text{brain}\right] \in \mathbb{R}^{(L + 2 + Q) \times D},
\end{equation}
where $\mathbf{c}_\text{start}, \mathbf{c}_\text{end} \in \mathbb{R}^{1 \times D}$ are the duration embeddings required by Stable Audio, and $L = 128$ is the length of the empty T5 sequence. Inside the processor, the last $Q$ tokens are split off
as $\mathbf{C}_\text{brain}$ and routed to the Adapter branch; the remaining $L + 2$ tokens are routed to the standard text branch. This keeps the DiT interface unchanged and makes the pipeline drop-in compatible with the Stable Audio denoising loop.

Adapting this mechanism to a transformer-based diffusion model requires addressing two architectural constraints. First, Stable Audio uses rotary positional embeddings (RoPE), which are applied to queries but not to conditioning keys. We preserve this asymmetry and avoid applying RoPE to $\mathbf{k}_\text{adap}$, as brain tokens do not possess a meaningful temporal ordering relative to the latent sequence. Second, the model employs grouped-query attention (GQA), in which the number of key-value heads $H_{kv}$ is smaller than the number of query heads $H_q$. To ensure compatibility, both text and brain branches replicate key-value heads via repeat interleaving to match the query head dimensionality before attention computation.

\subsection{Training Objective}

The DiT weights are frozen except for the cross-attention layers (keys, values, and layer norms of \textit{attn2}), which are fine-tuned jointly with the adapter. The VAE weights remain fully frozen throughout training. Ground-truth waveforms are encoded into latents $\mathbf{z}_0$ by the frozen VAE. At each training step, a random timestep $t$ is sampled, noise $\boldsymbol{\epsilon}$ is added to produce $\mathbf{z}_t$, and the model is trained to predict the v-prediction target $\mathbf{v}_t$ of Stable Audio's cosine noise schedule, using an MSE loss weighted by Min-SNR-$\gamma$ \citep{hang2023efficient}, which down-weights the contribution of low-SNR timesteps and stabilises training.

\subsection{Evaluation}

We evaluate reconstruction quality using four complementary metrics. Mel-SSIM and log-mel distance assess acoustic (spectral) fidelity. CLAP similarity measures cosine similarity in the CLAP embedding space~\citep{elizalde2023clap}, capturing high-level semantic correspondence between reconstructed and ground-truth audio. Pairwise identification accuracy follows the approach of~\cite{park2025natural}. 
We also compare our results with model baselines. The \emph{Brain-to-text} baseline predicts flan-T5 text embeddings (obtained with sound captions) from brain activity via ridge regression and conditions the DiT on the resulting estimated sequence, replicating the standard two-stage paradigm. \emph{Brain-only} passes $\mathbf{C}_\text{brain}$ directly as the full conditioning sequence in place of the T5 embeddings. \emph{Brain-plus-empty} adds $\mathbf{C}_\text{brain}$ as a scaled residual matched to the lambda Adapter on top of the empty-prompt T5 embedding (without the Adapter routing). \textcolor{black}{We also evaluate caption-based conditions. \emph{Caption-only} conditions Stable Audio on the ground-truth test caption without brain input; this oracle condition measures generation from semantic information unavailable to the brain decoder. \emph{Caption-plus-brain (zero-shot)} supplies the caption alongside brain tokens to the original MAAS model, which was trained with an empty text prompt.}

\textcolor{black}{Then we assess neural information over generative prior using two controls. First, MAAS is trained on 30 independently shuffled brain--stimulus assignments per condition, either unrestricted or within category, and evaluated by held-out pairwise identification. Second, ridge regression predicts audio embeddings from each ROI or all ROIs jointly, in a retrieval setting without audio generation \citep{ciferri2026r}. We report Top-1 and Top-5 retrieval among 72 held-out stimuli; RidgeCV selects alpha from [{0.01, 0.1, 1, 10, 100, 1000}] by internal cross-validation on the training set.}

Finally, to quantify the contribution of individual auditory ROIs, we introduce an ablation-based fidelity metric grounded in a computational model of auditory cortical processing.  Reconstructions are processed using the NSL Auditory-Cortical Toolbox \citep{chi2005multiresolution}, which models the transformation of sound along the auditory pathway from the cochlea to primary auditory cortex and produces a multiscale spectro-temporal cortical representation for details. For each sound, an auditory spectrogram was first estimated using 128 cochlear frequency channels, with centre frequencies defined relative to 440 Hz. Spectrotemporal modulation energy was then computed with using 32 temporal modulation rates logarithmically spaced between 1 and 128 Hz and 32 spectral modulation scales logarithmically spaced between 0.2 and 10 cycles/octave. Upward and downward temporal-rate channels were averaged over time and converted to absolute modulation energy, yielding for each sound a 3D representation of size 32 scales × 64 rates × 128 frequency channels (See Appendix \ref{par:ac_repre} for details). For each test sound, we compute the Pearson correlation $r$ between the flattened NSL representations of the reconstruction and of the ground truth. Correlations are averaged over the sounds of each category within each fold, and then across folds in Fisher-$z$ space ($\bar r = \tanh(\overline{\operatorname{artanh} r})$). The relative fidelity loss for ROI $i$ is $\Delta_i = (\bar r_{\text{rem}} - \bar r_{\text{all}})/\bar r_{\text{all}}$, where $\bar r_{\text{all}}$
and $\bar r_{\text{rem}}$ refer to reconstructions using all ROIs and with ROI $i$ removed or retained in isolation. Negative $\Delta_i$ indicates lower fidelity than the full model.

\section{Results}
\label{Results}

We evaluate MAAS on brain-to-audio reconstruction from fMRI. We first assess reconstruction performance against baselines, then examine qualitative reconstruction behaviour, and finally analyze the contribution of individual auditory cortical regions using our ablation-based fidelity metric.

\subsection{Quantitative Evaluation}
\label{sec:quantitative}

Table~\ref{tab:ablations} reports reconstruction performance on the held-out test set (72 sounds per fold), aggregated across four cross-validation folds and participants. 
Among brain-conditioning models, MAAS achieved the highest CLAP similarity (0.614), pairwise identification accuracy (85.4\%), and Mel-SSIM, while also yielding the lowest log-mel distance. Performance degraded progressively as components of the conditioning mechanism were removed, confirming the importance of both the adapter and backbone fine-tuning, despite routing through a semantically structured embedding space.

\textcolor{black}{Caption-only conditioning achieves high pairwise identification accuracy, as expected because the oracle caption explicitly provides semantic information. However, it performs substantially worse on both acoustic metrics than MAAS. This suggests that the caption and generative prior can produce a semantically compatible sound but do not recover the stimulus-specific spectrotemporal structure available in the neural responses. Adding captions only at inference to the original MAAS model constitutes an out-of-distribution condition because our model was trained exclusively with an empty prompt. Consistent with this, the zero-shot combination provides no systematic improvement over MAAS.}

\begin{table}[t]
\centering
\small
\caption{\textcolor{black}{Reconstruction performance. Higher is better for CLAP similarity, pairwise identification (ID) accuracy, and Mel-SSIM; lower is better for
log-mel distance. Caption-based conditions use ground-truth test captions.}}
\label{tab:ablations}
\begin{tabular}{lcccc}
\toprule
\textbf{Condition} &
\textbf{CLAP sim. $\uparrow$} &
\textbf{ID acc. $\uparrow$} &
\textbf{Mel-SSIM $\uparrow$} &
\textbf{Log-mel dist. $\downarrow$} \\
\midrule
\multicolumn{5}{l}{\textit{Brain-conditioning baselines}} \\
MAAS (our adapter) &
$0.614 \pm 0.13$ & $85.4 \pm 4.2\%$ &
$0.297 \pm 0.09$ & $0.152 \pm 0.04$ \\
MAAS (no backbone fine-tuning) &
$0.548 \pm 0.17$ & $73.1 \pm 4.8\%$ &
$0.283 \pm 0.08$ & $0.171 \pm 0.04$ \\
Brain-plus-empty &
$0.436 \pm 0.14$ & $63.1 \pm 5.7\%$ &
$0.273 \pm 0.10$ & $0.185 \pm 0.05$ \\
Brain-only &
$0.403 \pm 0.15$ & $58.3 \pm 5.9\%$ &
$0.226 \pm 0.08$ & $0.197 \pm 0.06$ \\
Two-stage fMRI-to-latent &
$0.509 \pm 0.16$ & $70.1 \pm 5.9\%$ &
$0.270 \pm 0.10$ & $0.177 \pm 0.08$ \\
\midrule
\multicolumn{5}{l}{\textit{Caption-based conditions}} \\
Caption-only &
$0.572 \pm 0.12$ & $87.9 \pm 5.4\%$ &
$0.166 \pm 0.08$ & $0.249 \pm 0.07$ \\
Caption-plus-brain (zero-shot) &
$0.610 \pm 0.14$ & $84.6 \pm 5.7\%$ &
$0.275 \pm 0.09$ & $0.167 \pm 0.06$ \\
Caption-plus-brain (retrained) &
$0.584 \pm 0.15$ & $82.4 \pm 5.1\%$ &
$0.311 \pm 0.11$ & $0.143 \pm 0.04$ \\
\bottomrule
\end{tabular}
\end{table}

\textcolor{black}{To address the possibility that reconstruction performance primarily reflects the generative prior, we performed two complementary controls (Table~\ref{tab:neural_controls}). Above-chance retrieval without audio generation demonstrates that stimulus information is decodable from the neural responses alone. The reduction in reconstruction accuracy after shuffling brain--stimulus pairings, including within semantic categories, further indicates that MAAS uses stimulus-specific neural information. Together, these controls support a neural contribution to
reconstruction that cannot be explained by the generative prior alone.}

\begin{table}[t]
\centering
\caption{Neural-information controls (mean $\pm$ SD, \%). Top: pairwise identification after
training with correct or shuffled brain--stimulus pairings (within-category or unrestricted stimulus identity
shuffled). Bottom: retrieval from fMRI without generation (72 test stimuli).}
\label{tab:neural_controls}
\small
\setlength{\tabcolsep}{3pt}
\begin{tabular*}{\linewidth}{@{}l@{\extracolsep{\fill}}cccc@{}}
\toprule
\textbf{Training pairing} & Correct & Within-category shuffle & Unrestricted shuffle & Chance \\
\midrule
Pairwise ID & $85.4 \pm 4.2$ & $73.6 \pm 4.0$ & $53.4 \pm 2.7$ & $50.0$ \\
\end{tabular*}

\medskip
\begin{tabular*}{\linewidth}{@{}l@{\extracolsep{\fill}}cccccccc@{}}
\toprule
\textbf{Retrieval} & HG & PT & PP & mSTG & pSTG & aSTG & All ROIs & Chance \\
\midrule
Top-1 & $12.5 \pm 2.9$ & $14.2 \pm 3.7$ & $8.3 \pm 2.9$ & $12.2 \pm 3.3$ & $10.1 \pm 2.8$ & $4.9 \pm 1.6$ & $17.5 \pm 3.6$ & $1.39$ \\
Top-5 & $44.8 \pm 4.9$ & $47.9 \pm 4.7$ & $39.2 \pm 3.8$ & $43.4 \pm 4.3$ & $39.9 \pm 3.8$ & $26.4 \pm 3.1$ & $56.4 \pm 4.8$ & $6.94$ \\
\bottomrule
\end{tabular*}
\end{table}

\subsection{Qualitative Reconstruction}
MAAS produced high-quality sound-reconstructions from fMRI responses. We provide audio examples in the following site: \url{https://fidelioc55.github.io/brainSounds-website/}.

\begin{figure}[h]
    \centering
    \includegraphics[width=\linewidth]{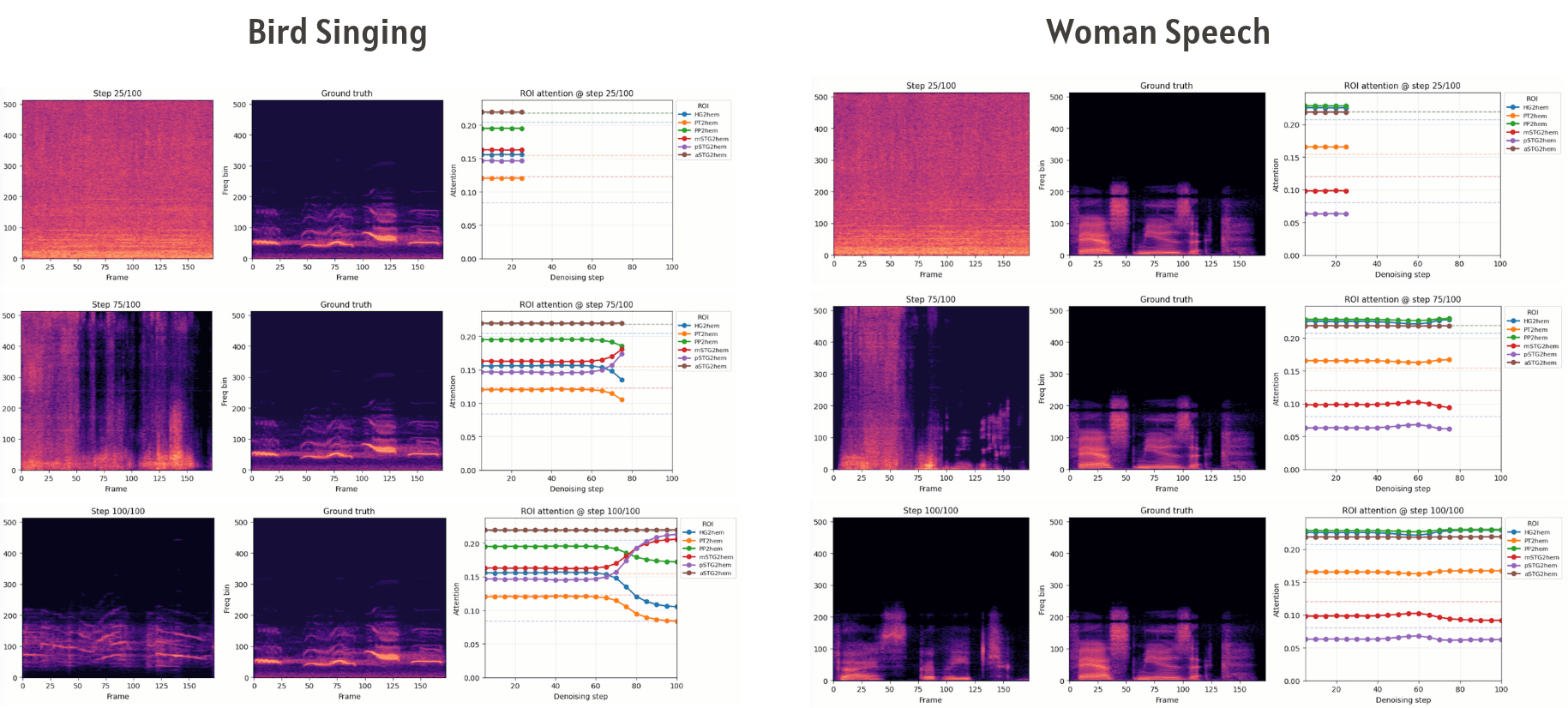}
    \caption{Analysis of cortical contributions in MAAS.
    Evolution of per-ROI transported attention $\bar{\alpha}_t(r)$ across 100 denoising steps for two test samples. Each row shows the
    reconstructed spectrogram at a given denoising step alongside the ground truth and the corresponding ROI attention profile.}
    \label{fig:denoising_steps}
\end{figure}

Figure~\ref{fig:denoising_steps} illustrates the denoising trajectory for two representative test samples. The final reconstructions capture key spectrotemporal characteristics of the ground-truth signals, including harmonic structure in speech and vocalizations and broadband dynamics in environmental sounds. Intermediate denoising steps reveal a structured progression: early steps capture coarse temporal envelopes, while later stages refine spectral detail. This behavior is consistent with the coarse-to-fine nature of diffusion-based generation ~\citep{ho2020denoising}.

We further analyzed attention weights during inference. The evolution of ROI-specific attention across denoising steps shows that different cortical regions contribute at different stages of synthesis. In particular, a marked reweighting occurs around steps 60–70, corresponding to the transition from low to high signal-to-noise ratio in the cosine schedule ~\citep{nichol2021improved}. This pattern suggests that the adapter dynamically changes how it routes brain-derived information as generation progresses. See Appendix~\ref{par:attn_decomp} for a full description of attention decomposition.

Importantly, the attention profiles were strongly sound-dependent: different stimuli elicit distinct patterns of ROI weighting throughout the denoising process. This indicates that the contribution of individual cortical regions is not fixed, but adapts to the specific acoustic content of each sound. This observation is consistent with the ROI-level fidelity analysis in Section \ref{sec:cortic_contrib}, where different regions contribute differentially across sound categories.


\subsection{Cortical Contribution Analysis}
\label{sec:cortic_contrib}

\begin{figure}[h]
    \centering
    \includegraphics[width=.99\linewidth]{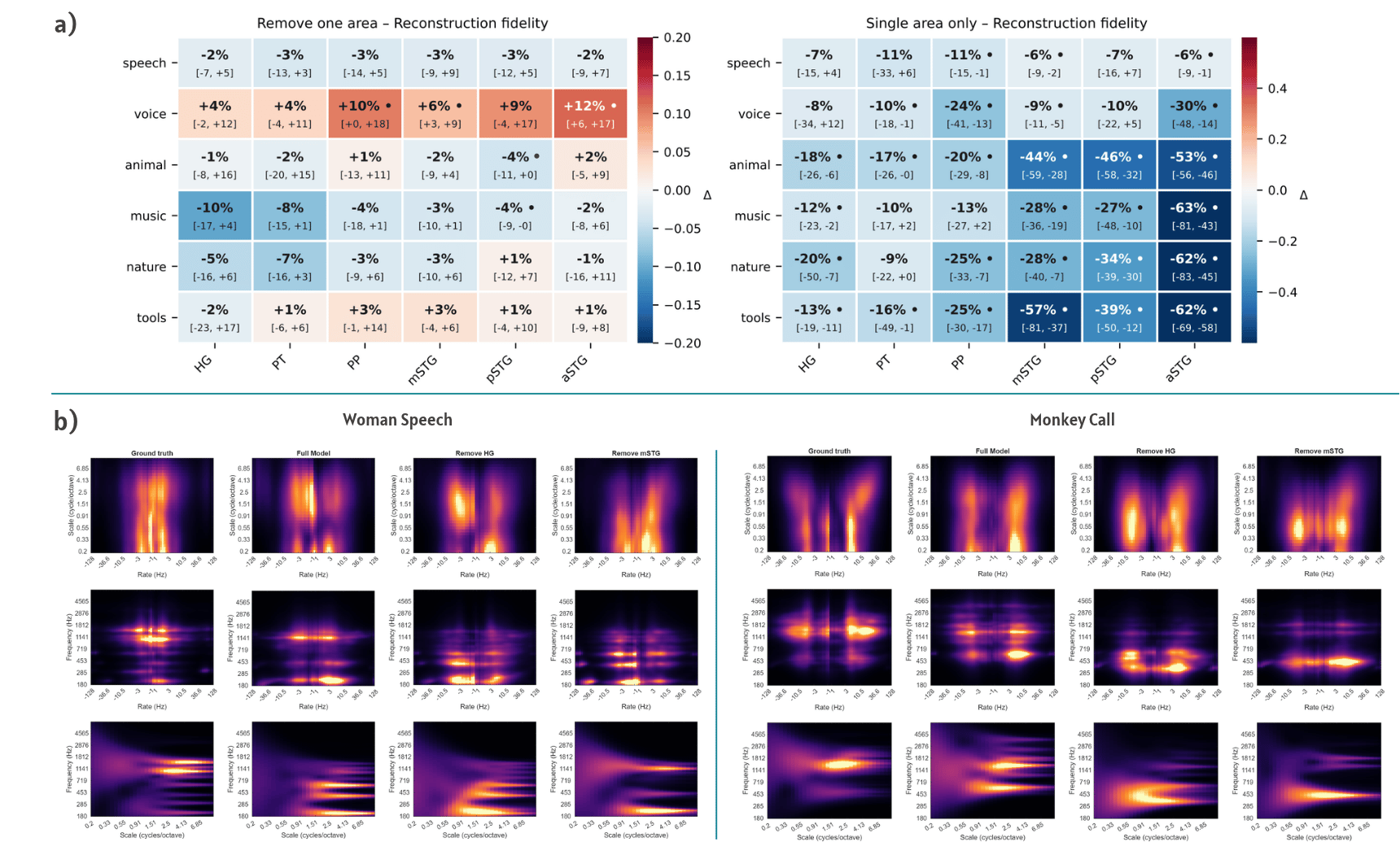}
    \caption{\textbf{(a)} Relative change in NSL-based reconstruction fidelity, $\Delta_i$, by ROI and semantic category, computed from Fisher-$z$ averaged correlations across the four cross-validation folds; brackets show
    the range [min, max] of fold-wise $\Delta_i$ (\%). Dots mark cells in which all four folds agree in
    sign. Left: leave-one-ROI-out analysis. Right: single-ROI analysis. Each partial-ROI condition uses a
    separately trained model with the same fixed mask during training and testing; negative values
    indicate lower fidelity.
    \textbf{(b)} NSL-derived spectrotemporal representations for two
    individual stimuli, woman speech and monkey call. Columns compare
    the ground truth, all-ROI model, and models trained without HG or
    mSTG. Rows show scale--rate, frequency--rate, and frequency--scale
    projections, respectively, obtained by averaging over the remaining
    dimension of the time-averaged representation.}
    \label{fig:roi_ablations}
\end{figure}

\textcolor{black}{We examined the contribution of auditory ROIs using condition-specific models trained and evaluated
with identical fixed ROI masks in all four folds (Figure~\ref{fig:roi_ablations}a). Removing a single ROI produced
comparatively modest changes in fidelity (averaged across categories, between $-2.8\%$ and $+1.7\%$
per ROI), whereas models conditioned on a single ROI lost up to $83\%$ of the full-model fidelity in
individual folds (removal vs.\ single-ROI, $p<0.001$; repeated-measures ANOVA across the four folds
on Fisher-$z$ differences from the all-ROI model). Thus, for the
reconstruction model, information across ROIs is largely redundant, and the remaining regions can
typically compensate for an omitted one. Within this smaller range, the leave-one-ROI-out analysis
showed an effect of ROI ($p=0.044$) whose ordering mirrored the single-ROI analysis:
removing HG or PT produced the largest losses on average ($-2.8\%$ and $-2.5\%$), whereas removing
STG regions left fidelity close to the all-ROI reference ($-0.3\%$ to $+1.7\%$). The cost of removal
was most evident for music, where every removal reduced fidelity on average and removing HG and PT
produced the largest losses ($-10.4\%$ and $-8.4\%$), and for nature sounds (HG: $-5.3\%$, PT:
$-7.0\%$).
Single-ROI fidelity varied systematically across regions (effect of ROI, $p<0.001$). Early supratemporal-plane
ROIs retained most of the fidelity in isolation (PT: $-12.2\%$, HG: $-13.1\%$, PP: $-19.7\%$), whereas
lateral STG regions retained less (pSTG: $-26.9\%$, mSTG: $-28.4\%$, aSTG: $-45.9\%$; supratemporal
vs.\ STG, $p=0.003$, paired $t$-test across folds), with the largest loss in the most anterior region
(posterior-to-anterior trend, $p=0.015$, $t$-test on the fold-wise slopes). Together with the
leave-one-out ordering, this pattern is consistent with spectrotemporal detail being most directly
available in early auditory areas and with increasingly abstract representations along the STG
\citep{norman2018neural,giordano2023intermediate}. The effect of ROI also varied across
categories, and we describe these category-specific patterns qualitatively. For speech, every
removal reduced fidelity on average ($-2\%$ to $-3\%$) and all single-ROI models, including aSTG,
performed similarly ($-6\%$ to $-11\%$), consistent with distributed encoding of speech across the
STG \citep{hamilton2021parallel}. For non-speech vocalizations, removing any ROI slightly increased
fidelity on average ($+4\%$ to $+12\%$), indicating that additional inputs do not invariably improve
reconstruction under this metric.
Finally, the individual examples in Figure~\ref{fig:roi_ablations}b complement
these quantitative results by illustrating how different ROI
conditions alter the reconstructed spectrotemporal structure.}

\section{Discussion}

MAAS demonstrates that direct conditioning of a latent audio diffusion transformer on fMRI signals is both feasible and advantageous. The consistent performance gains across semantic and perceptual metrics suggest that bypassing intermediate feature prediction preserves information that is otherwise lost in the regression bottleneck, a finding that aligns with recent evidence in the visual domain~\citep{feng2025towards} and extends it to the architecturally distinct setting of DiT-based audio generation.

\textcolor{black}{The additional controls suggest that neural conditioning contributes to resolving the acoustic ambiguity inherent in semantic descriptions. A caption can identify a sound source while remaining compatible with many different acoustic realizations. The higher acoustic fidelity obtained with brain conditioning, including when oracle captions are available, suggests that fMRI responses provide constraints that help the generator recover aspects of the particular sound that was heard. The performance reduction under within-category shuffling further supports this interpretation: preserving category correspondence alone does not reproduce the performance obtained with correctly paired neural responses. Together, these findings suggest that MAAS uses neural information beyond broad semantic compatibility to guide the acoustic realization of generated sounds. The trade-off observed under joint caption and brain conditioning also indicates that semantic recognizability and acoustic fidelity capture distinct aspects of reconstruction.}

From a technical standpoint, the adapter design addresses a non-trivial compatibility problem: injecting continuous, non-linguistic representations into a model built exclusively around text conditioning, while preserving RoPE asymmetry and GQA structure. The fact that fine-tuning only the cross-attention layers is necessary to achieve strong performance suggests that the generative prior of Stable Audio is preserved, with the brain pathway acting as a targeted guidance. 

The ablation analysis reveals a hierarchical behaviour that mirrors what is known about the functional organization of the human auditory cortex \citep{kell2018task,moerel2014anatomical,rolls2023auditory}: early areas (HG, PT) carry the least redundant and most self-sufficient spectrotemporal information, while higher-order STG regions are largely redundant with them and individually insufficient for faithful reconstruction of non-vocal sounds. This is consistent with evidence that primary auditory cortex responses are largely explained by spectrotemporal modulations ~\citep{santoro2014encoding,santoro2017reconstructing}, whereas non-primary regions encode higher-order structure that goes beyond what modulation statistics can capture ~\citep{norman2018neural, landemard2021distinct}, representations that are categorically informative but perceptually incomplete without the low-level substrate provided by earlier areas. The fact that higher-order regions alone produce the largest fidelity drops suggests that their intermediate acoustic-to-semantic representations ~\citep{giordano2023intermediate}, while sufficient for recognition, do not carry enough fine-grained spectro-temporal detail to constrain the generative process toward perceptually faithful synthesis.

Several limitations should be acknowledged. The dataset comprises only five participants and 288 stimuli of 1 second duration, which constrains the generalizability of the reconstruction results. fMRI carries inherent limitations in temporal resolution: the BOLD signal integrates neural activity over several seconds and, with clustered acquisition and GLM-based beta estimation, the input to MAAS reflects a temporally blurred snapshot of auditory cortical processing. That the model achieves the reported reconstruction quality under these constraints is encouraging, and suggests that the spatial pattern of activation across ROIs carries substantial generative information even in the absence of fine-grained temporal dynamics, a hypothesis that could be tested with higher-temporal-resolution modalities. While the attention decomposition provides a mechanistic window into the generative process, the relationship between attention routing and ablation outcomes suggests that the adapter's internal representations warrant further investigation.

\section{Conclusion}

We presented Multi-Area Auditory Synthesis (MAAS), a framework for directly conditioning a latent audio diffusion transformer on fMRI signals. The proposed adapter provides a mechanism for integrating neural representations into DiT-based generative models.
Beyond reconstruction performance, MAAS offers a principled framework for probing the functional contribution of individual auditory cortical regions to sound perception. By conditioning an end-to-end generative model on structured brain representations, our approach enables a generative perspective on cortical information content, complementing traditional encoding and decoding analyses.

Looking forward, the direct conditioning paradigm opens several promising directions. In the clinical domain, the ability to reconstruct perceived audio from cortical activity could inform the development of auditory brain-computer interfaces for patients with severe motor impairments or disorders of consciousness, where passive listening paradigms are the only viable option.

\section*{Acknowledgements}
This work was funded by the Dutch Research Council (NWO Grant 406.20.GO.030) and the European Research Council through ERC-2024-SyG NASCE (grant agreement 101167313). This work was also supported by NEXTGENERATIONEU (NGEU) and funded by the Italian Ministry of University and Research (MUR), National Recovery and Resilience Plan (NRRP), project MNESYS (PE0000006) (to NT)– A Multiscale integrated approach to the study of the nervous system in health and disease (DN. 1553 11.10.2022); by the MUR-PNRR M4C2I1.3 PE6 project PE00000019 Heal Italia (to NT).

\section*{Competing Interests}
M.F. is employed by Tether Evo. The company had no role in the design, execution, or reporting of this work. The other authors declare no competing interests.

\bibliography{neurips_2026}

@article{hamilton2021parallel,
  title={Parallel and distributed encoding of speech across human auditory cortex},
  author={Hamilton, Liberty S and Oganian, Yulia and Hall, Jeffery and Chang, Edward F},
  journal={Cell},
  volume={184},
  number={18},
  pages={4626--4639},
  year={2021},
  publisher={Elsevier}
}

@article{ciferri2026r,
  title={R\&B-rhythm and brain: Cross-subject decoding of music from human brain activity},
  author={Ciferri, Matteo and Ferrante, Matteo and Toschi, Nicola},
  journal={Neural Networks},
  pages={109195},
  year={2026},
  publisher={Elsevier}
}

@article{esteban2019fmriprep,
  title={fMRIPrep: a robust preprocessing pipeline for functional MRI},
  author={Esteban, Oscar and Markiewicz, Christopher J and Blair, Ross W and Moodie, Craig A and Isik, A Ilkay and Erramuzpe, Asier and Kent, James D and Goncalves, Mathias and DuPre, Elizabeth and Snyder, Madeleine and others},
  journal={Nature methods},
  volume={16},
  number={1},
  pages={111--116},
  year={2019},
  publisher={Nature Publishing Group US New York}
}

@article{musecontrollite,
  title={Musecontrollite: Multifunctional music generation with lightweight conditioners},
  author={Tsai, Fang-Duo and Wu, Shih-Lun and Lee, Weijaw and Yang, Sheng-Ping and Chen, Bo-Rui and Cheng, Hao-Chung and Yang, Yi-Hsuan},
  journal={arXiv preprint arXiv:2506.18729},
  year={2025}
}

@article{liu2024audioldm,
  title={Audioldm 2: Learning holistic audio generation with self-supervised pretraining},
  author={Liu, Haohe and Yuan, Yi and Liu, Xubo and Mei, Xinhao and Kong, Qiuqiang and Tian, Qiao and Wang, Yuping and Wang, Wenwu and Wang, Yuxuan and Plumbley, Mark D},
  journal={IEEE/ACM Transactions on Audio, Speech, and Language Processing},
  volume={32},
  pages={2871--2883},
  year={2024},
  publisher={IEEE}
}

@article{tsai2024audio,
  title={Audio prompt adapter: Unleashing music editing abilities for text-to-music with lightweight finetuning},
  author={Tsai, Fang-Duo and Wu, Shih-Lun and Kim, Haven and Chen, Bo-Yu and Cheng, Hao-Chung and Yang, Yi-Hsuan},
  journal={arXiv preprint arXiv:2407.16564},
  year={2024}
}

@article{kowalski1996analysis,
  title={Analysis of dynamic spectra in ferret primary auditory cortex. I. Characteristics of single-unit responses to moving ripple spectra},
  author={Kowalski, Nina and Depireux, Didier A and Shamma, Shihab A},
  journal={Journal of neurophysiology},
  volume={76},
  number={5},
  pages={3503--3523},
  year={1996}
}

@article{rolls2023auditory,
  title={Auditory cortical connectivity in humans},
  author={Rolls, Edmund T and Rauschecker, Josef P and Deco, Gustavo and Huang, Chu-Chung and Feng, Jianfeng},
  journal={Cerebral Cortex},
  volume={33},
  number={10},
  pages={6207--6227},
  year={2023},
  publisher={Oxford University Press}
}

@article{moerel2014anatomical,
  title={An anatomical and functional topography of human auditory cortical areas},
  author={Moerel, Michelle and De Martino, Federico and Formisano, Elia},
  journal={Frontiers in neuroscience},
  volume={8},
  pages={225},
  year={2014},
  publisher={Frontiers Media SA}
}

@inproceedings{rombach2022high,
  title={High-resolution image synthesis with latent diffusion models},
  author={Rombach, Robin and Blattmann, Andreas and Lorenz, Dominik and Esser, Patrick and Ommer, Bj{\"o}rn},
  booktitle={Proceedings of the IEEE/CVF conference on computer vision and pattern recognition},
  pages={10684--10695},
  year={2022}
}

@article{landemard2021distinct,
  title={Distinct higher-order representations of natural sounds in human and ferret auditory cortex},
  author={Landemard, Agn{\`e}s and Bimbard, C{\'e}lian and Demen{\'e}, Charlie and Shamma, Shihab and Norman-Haignere, Sam and Boubenec, Yves},
  journal={Elife},
  volume={10},
  pages={e65566},
  year={2021},
  publisher={eLife Sciences Publications, Ltd}
}

@article{norman2018neural,
  title={Neural responses to natural and model-matched stimuli reveal distinct computations in primary and nonprimary auditory cortex},
  author={Norman-Haignere, Sam V and McDermott, Josh H},
  journal={PLoS biology},
  volume={16},
  number={12},
  pages={e2005127},
  year={2018},
  publisher={Public Library of Science San Francisco, CA USA}
}

@article{santoro2014encoding,
  title={Encoding of natural sounds at multiple spectral and temporal resolutions in the human auditory cortex},
  author={Santoro, Roberta and Moerel, Michelle and De Martino, Federico and Goebel, Rainer and Ugurbil, Kamil and Yacoub, Essa and Formisano, Elia},
  journal={PLoS computational biology},
  volume={10},
  number={1},
  pages={e1003412},
  year={2014},
  publisher={Public Library of Science San Francisco, USA}
}

@article{chi2005multiresolution,
  title={Multiresolution spectrotemporal analysis of complex sounds},
  author={Chi, Taishih and Ru, Powen and Shamma, Shihab A},
  journal={The Journal of the Acoustical Society of America},
  volume={118},
  number={2},
  pages={887--906},
  year={2005},
  publisher={AIP Publishing}
}

@inproceedings{elizalde2023clap,
  title={Clap learning audio concepts from natural language supervision},
  author={Elizalde, Benjamin and Deshmukh, Soham and Al Ismail, Mahmoud and Wang, Huaming},
  booktitle={ICASSP 2023-2023 IEEE International Conference on Acoustics, Speech and Signal Processing (ICASSP)},
  pages={1--5},
  year={2023},
  organization={IEEE}
}

@article{ho2020denoising,
  title={Denoising diffusion probabilistic models},
  author={Ho, Jonathan and Jain, Ajay and Abbeel, Pieter},
  journal={Advances in Neural Information Processing Systems},
  volume={33},
  pages={6840--6851},
  year={2020}
}

@inproceedings{nichol2021improved,
  title={Improved denoising diffusion probabilistic models},
  author={Nichol, Alexander Quinn and Dhariwal, Prafulla},
  booktitle={International Conference on Machine Learning},
  pages={8162--8171},
  year={2021},
  organization={PMLR}
}

@inproceedings{hershey2017cnn,
  title={CNN architectures for large-scale audio classification},
  author={Hershey, Shawn and Chaudhuri, Sourish and Ellis, Daniel PW and Gemmeke, Jort F and Jansen, Aren and Moore, R Channing and Plakal, Manoj and Platt, Devin and Saurous, Rif A and Seybold, Bryan and others},
  booktitle={2017 ieee international conference on acoustics, speech and signal processing (icassp)},
  pages={131--135},
  year={2017},
  organization={IEEE}
}

@article{chung2024scaling,
  title={Scaling instruction-finetuned language models},
  author={Chung, Hyung Won and Hou, Le and Longpre, Shayne and Zoph, Barret and Tay, Yi and Fedus, William and Li, Yunxuan and Wang, Xuezhi and Dehghani, Mostafa and Brahma, Siddhartha and others},
  journal={Journal of Machine Learning Research},
  volume={25},
  number={70},
  pages={1--53},
  year={2024}
}

@inproceedings{hang2023efficient,
  title={Efficient diffusion training via min-snr weighting strategy},
  author={Hang, Tiankai and Gu, Shuyang and Li, Chen and Bao, Jianmin and Chen, Dong and Hu, Han and Geng, Xin and Guo, Baining},
  booktitle={Proceedings of the IEEE/CVF international conference on computer vision},
  pages={7441--7451},
  year={2023}
}

@inproceedings{ainslie2023gqa,
  title={Gqa: Training generalized multi-query transformer models from multi-head checkpoints},
  author={Ainslie, Joshua and Lee-Thorp, James and De Jong, Michiel and Zemlyanskiy, Yury and Lebr{\'o}n, Federico and Sanghai, Sumit},
  booktitle={Proceedings of the 2023 Conference on Empirical Methods in Natural Language Processing},
  pages={4895--4901},
  year={2023}
}

@article{su2024roformer,
  title={Roformer: Enhanced transformer with rotary position embedding},
  author={Su, Jianlin and Ahmed, Murtadha and Lu, Yu and Pan, Shengfeng and Bo, Wen and Liu, Yunfeng},
  journal={Neurocomputing},
  volume={568},
  pages={127063},
  year={2024},
  publisher={Elsevier}
}

@article{kay2013glmdenoise,
  title={GLMdenoise: a fast, automated technique for denoising task-based fMRI data},
  author={Kay, Kendrick and Rokem, Ariel and Winawer, Jonathan and Dougherty, Robert and Wandell, Brian},
  journal={Frontiers in neuroscience},
  volume={7},
  pages={54426},
  year={2013},
  publisher={Frontiers}
}

@article{ye2023ip,
  title={Ip-adapter: Text compatible image prompt adapter for text-to-image diffusion models},
  author={Ye, Hu and Zhang, Jun and Liu, Sibo and Han, Xiao and Yang, Wei},
  journal={arXiv preprint arXiv:2308.06721},
  year={2023}
}

@article{kell2018task,
  title={A task-optimized neural network replicates human auditory behavior, predicts brain responses, and reveals a cortical processing hierarchy},
  author={Kell, Alexander JE and Yamins, Daniel LK and Shook, Erica N and Norman-Haignere, Sam V and McDermott, Josh H},
  journal={Neuron},
  volume={98},
  number={3},
  pages={630--644},
  year={2018},
  publisher={Elsevier}
}

@inproceedings{radford2021learning,
  title={Learning transferable visual models from natural language supervision},
  author={Radford, Alec and Kim, Jong Wook and Hallacy, Chris and Ramesh, Aditya and Goh, Gabriel and Agarwal, Sandhini and Sastry, Girish and Askell, Amanda and Mishkin, Pamela and Clark, Jack and others},
  booktitle={International conference on machine learning},
  pages={8748--8763},
  year={2021},
  organization={PmLR}
}

@article{oquab2023dinov2,
  title={Dinov2: Learning robust visual features without supervision},
  author={Oquab, Maxime and Darcet, Timoth{\'e}e and Moutakanni, Th{\'e}o and Vo, Huy and Szafraniec, Marc and Khalidov, Vasil and Fernandez, Pierre and Haziza, Daniel and Massa, Francisco and El-Nouby, Alaaeldin and others},
  journal={arXiv preprint arXiv:2304.07193},
  year={2023}
}

@article{ye2025generative,
  title={Generative language reconstruction from brain recordings},
  author={Ye, Ziyi and Ai, Qingyao and Liu, Yiqun and de Rijke, Maarten and Zhang, Min and Lioma, Christina and Ruotsalo, Tuukka},
  journal={Communications Biology},
  volume={8},
  number={1},
  pages={346},
  year={2025},
  publisher={Nature Publishing Group UK London}
}

@article{chen2023cinematic,
  title={Cinematic mindscapes: High-quality video reconstruction from brain activity},
  author={Chen, Zijiao and Qing, Jiaxin and Zhou, Juan Helen},
  journal={Advances in Neural Information Processing Systems},
  volume={36},
  pages={24841--24858},
  year={2023}
}

@inproceedings{evans2025stable,
  title={Stable audio open},
  author={Evans, Zach and Parker, Julian D and Carr, CJ and Zukowski, Zack and Taylor, Josiah and Pons, Jordi},
  booktitle={ICASSP 2025-2025 IEEE International Conference on Acoustics, Speech and Signal Processing (ICASSP)},
  pages={1--5},
  year={2025},
  organization={IEEE}
}

@article{giordano2023intermediate,
  title={Intermediate acoustic-to-semantic representations link behavioral and neural responses to natural sounds},
  author={Giordano, Bruno L and Esposito, Michele and Valente, Giancarlo and Formisano, Elia},
  journal={Nature Neuroscience},
  volume={26},
  number={4},
  pages={664--672},
  year={2023},
  publisher={Nature Publishing Group US New York}
}

@article{denk2025text,
  title={Text-to-music generation models capture musical semantic representations in the human brain},
  author={Denk, Timo I and Takagi, Yu and Matsuyama, Takuya and Agostinelli, Andrea and Nakai, Tomoya and Frank, Christian and Nishimoto, Shinji},
  journal={Nature Communications},
  year={2025},
  publisher={Nature Publishing Group UK London}
}

@article{park2025natural,
  title={Natural sounds can be reconstructed from human neuroimaging data using deep neural network representation},
  author={Park, Jong-Yun and Tsukamoto, Mitsuaki and Tanaka, Misato and Kamitani, Yukiyasu},
  journal={PLoS biology},
  volume={23},
  number={7},
  pages={e3003293},
  year={2025},
  publisher={Public Library of Science San Francisco, CA USA}
}

@article{santoro2017reconstructing,
  title={Reconstructing the spectrotemporal modulations of real-life sounds from fMRI response patterns},
  author={Santoro, Roberta and Moerel, Michelle and De Martino, Federico and Valente, Giancarlo and Ugurbil, Kamil and Yacoub, Essa and Formisano, Elia},
  journal={Proceedings of the National Academy of Sciences},
  volume={114},
  number={18},
  pages={4799--4804},
  year={2017},
  publisher={National Academy of Sciences}
}

@inproceedings{feng2025towards,
    title={Towards Interpretable Visual Decoding with Attention to Brain Representations},
    author={Pinyuan Feng and Hossein Adeli and Wenxuan Guo and Fan Cheng and Ethan Hwang and Nikolaus Kriegeskorte},
    booktitle={The Fourteenth International Conference on Learning Representations},
    year={2026},
    url={https://openreview.net/forum?id=YWlYITAhMC}
}

@article{shirakawa2025spurious,
  title={Spurious reconstruction from brain activity},
  author={Shirakawa, Ken and Nagano, Yoshihiro and Tanaka, Misato and Aoki, Shuntaro C and Muraki, Yusuke and Majima, Kei and Kamitani, Yukiyasu},
  journal={Neural Networks},
  volume={190},
  pages={107515},
  year={2025},
  publisher={Elsevier}
}

@inproceedings{takagi2023high,
  title={High-resolution image reconstruction with latent diffusion models from human brain activity},
  author={Takagi, Yu and Nishimoto, Shinji},
  booktitle={Proceedings of the IEEE/CVF conference on computer vision and pattern recognition},
  pages={14453--14463},
  year={2023}
}

@article{scotti2024mindeye2,
  title={Mindeye2: Shared-subject models enable fmri-to-image with 1 hour of data},
  author={Scotti, Paul S and Tripathy, Mihir and Villanueva, Cesar Kadir Torrico and Kneeland, Reese and Chen, Tong and Narang, Ashutosh and Santhirasegaran, Charan and Xu, Jonathan and Naselaris, Thomas and Norman, Kenneth A and others},
  journal={arXiv preprint arXiv:2403.11207},
  year={2024}
}

@article{luo2023brain,
  title={Brain diffusion for visual exploration: Cortical discovery using large scale generative models},
  author={Luo, Andrew and Henderson, Maggie and Wehbe, Leila and Tarr, Michael},
  journal={Advances in Neural Information Processing Systems},
  volume={36},
  pages={75740--75781},
  year={2023}
}

@article{ferrante2024decoding,
  title={Decoding visual brain representations from electroencephalography through knowledge distillation and latent diffusion models},
  author={Ferrante, Matteo and Boccato, Tommaso and Bargione, Stefano and Toschi, Nicola},
  journal={Computers in Biology and Medicine},
  volume={178},
  pages={108701},
  year={2024},
  publisher={Elsevier}
}

@article{ciferri2025reconstructing,
  title={Reconstructing music perception from brain activity using a prior guided diffusion model},
  author={Ciferri, Matteo and Ferrante, Matteo and Toschi, Nicola},
  journal={Scientific Reports},
  volume={15},
  number={1},
  pages={42108},
  year={2025},
  publisher={Nature Publishing Group UK London}
}

@Article{Ozcelik2023,
author={Ozcelik, Furkan
and VanRullen, Rufin},
title={Natural scene reconstruction from fMRI signals using generative latent diffusion},
journal={Scientific Reports},
year={2023},
month={Sep},
day={20},
volume={13},
number={1},
pages={15666},
issn={2045-2322},
doi={10.1038/s41598-023-42891-8},
url={https://doi.org/10.1038/s41598-023-42891-8}
}

@article{ferrante2023eyes,
    author = {Ferrante, Matteo and Boccato, Tommaso and Ozcelik, Furkan and VanRullen, Rufin and Toschi, Nicola},
    title = "{Through their eyes: Multi-subject brain decoding with simple alignment techniques}",
    journal = {Imaging Neuroscience},
    volume = {2},
    pages = {1-21},
    year = {2024},
    month = {05},
    issn = {2837-6056},
    doi = {10.1162/imag\_a\_00170},
    url = {https://doi.org/10.1162/imag\_a\_00170},
}

@article {glmsingle,
article_type = {journal},
title = {Improving the accuracy of single-trial fMRI response estimates using GLMsingle},
author = {Prince, Jacob S and Charest, Ian and Kurzawski, Jan W and Pyles, John A and Tarr, Michael J and Kay, Kendrick N},
editor = {Kok, Peter and de Lange, Floris P and Kok, Peter and Turner, Benjamin},
volume = 11,
year = 2022,
month = {nov},
pub_date = {2022-11-29},
pages = {e77599},
citation = {eLife 2022;11:e77599},
doi = {10.7554/eLife.77599},
url = {https://doi.org/10.7554/eLife.77599},
journal = {eLife},
issn = {2050-084X},
publisher = {eLife Sciences Publications, Ltd},
}
\bibliographystyle{plainnat}


\appendix
\setcounter{table}{0}
\renewcommand{\thetable}{A\arabic{table}}
\setcounter{figure}{0}
\renewcommand{\thefigure}{A\arabic{figure}}

\section{Technical appendices and supplementary material}

We provide code for the reproducibility of our pipeline at the following link: \url{https://anonymous.4open.science/r/brainSounds_nips-DFC1}.

\paragraph{Attention decomposition.}
\label{par:attn_decomp}

MAAS exposes two distinct loci of attention over auditory ROIs, which can be composed to obtain a single ROI-level influence map. Let $M(k, r)$ denote the cross-attention weight assigned by the Brain Token Generator from query token $k$ to ROI $r$, and let $B_t(l, k)$ denote the attention weight assigned by the fMRI Adapter branch from latent position $l$ to brain token $k$ at denoising timestep $t$. The net influence of ROI $r$ on latent position $l$ is obtained by composing the two maps:
\begin{equation}
    \mathrm{transport}_t(l, r) = \sum_k B_t(l, k)\cdot M(k, r).
    \label{eq:transport}
\end{equation}
Averaging over latent positions yields a per-ROI influence score
$\bar{\alpha}_t(r) = \frac{1}{L}\sum_l \mathrm{transport}_t(l,r)$.

\paragraph{Hyperparameter selection.}
\label{par:Hyperparameter}

We performed a grid search over key hyperparameters on the validation split.  The explored ranges and final selected configuration are summarized in Table~\ref{tab:hyperparams}. To enable classifier-free guidance at inference time, we apply conditioning dropout with probability $p = 0.05$ by zeroing $\mathbf{C}_\text{brain}$ during training. The model is optimised with AdamW ($\text{lr} = 10^{-4}$, weight decay $= 10^{-4}$), cosine annealing to $\text{lr}_\text{min} = 10^{-6}$ and gradient clipping at norm 1.0. Experiments were run on a shared server equipped with eight NVIDIA A100 GPUs (80 GB each), 256 CPU threads, and 2 TB of system memory. Each training run used a single GPU for 60 epochs, and the main configuration updated 181M trainable parameters.

\begin{table}[h]
\centering
\small
\begin{tabular}{lcc}
\toprule
\textbf{Hyperparameter} & \textbf{Selected} & \textbf{Search range} \\
\midrule
Number of brain queries $Q$ & 16 & \{6, 16, 64, 128\} \\
Conditioning strategy & adapter & \{brain-only, residual, adapter\} \\
BrainTokenGen. layers & 1 & \{1, 2, 4\} \\
Transformer dimension & 1024 & \{512, 1024, 2048\} \\
Learning rate & $1\times10^{-4}$ & $\{1\cdot10^{-5}, 5\cdot10^{-4}, 1\cdot10^{-4}\}$ \\
Weight decay & $1\times10^{-4}$ & $\{0, 1\cdot10^{-5}, 1\cdot10^{-4}, 1\cdot10^{-3}\}$ \\
Effective batch size & 64 & \{16, 32, 64\} \\
Min-SNR $\gamma$ & 3.0 & \{0, 1, 3, 5\} \\
CFG dropout & 0.05 & \{0, 0.05, 0.1, 0.3\} \\
Guidance scale & 3.0 & \{1.0, 2.0, 3.0, 5.0\} \\
\bottomrule
\end{tabular}
\vspace{0.6em}
\caption{Hyperparameter selection. We report the final configuration and the grid explored during validation.}
\label{tab:hyperparams}
\end{table}

\paragraph{External validation on an independent 3T dataset.}

\textcolor{black}{To assess whether the applicability of MAAS extends beyond the original 7T acquisition, we evaluated it on the public 3T dataset of \cite{park2025natural}, comprising five independently acquired participants and natural sounds of 8 or 10 seconds. This experiment introduces variation in field strength, stimulus duration, participants, and acquisition protocol, providing an external assessment of the brain-conditioning approach under different measurement conditions. We considered two strategies for estimating stimulus-related fMRI responses. The first followed the delayed-volume averaging procedure of the original study: BOLD volumes were temporally aligned with the stimuli by accounting for the hemodynamic delay (2 seconds), divided into overlapping windows, and then averaged to obtain stimulus-related response patterns. For the second strategy, we estimated stimulus-evoked beta coefficients via GLMsingle~\citep{glmsingle}. Unlike delayed-volume averaging, this approach models the BOLD time series using the experimental design to estimate response amplitudes associated with stimulus presentations. In both strategies we used fMRIprep library \citep{esteban2019fmriprep} to process neural data. The resulting estimates were organized into the same six auditory ROI inputs used by MAAS (HG, PT, PP, mSTG, pSTG, and aSTG). Using stimulus-wise beta representations improves comparability with the neural input format of our 7T experiments, while differences in acquisition
and response-estimation procedures remain.}

\textcolor{black}{MAAS achieved pairwise identification accuracies of
$83.9 \pm 3.8\%$ with delayed-volume averaging and $86.1 \pm 3.5\%$ with GLMsingle. The higher point estimate with GLMsingle suggests that response estimation can affect reconstruction quality, while the performance obtained with both strategies supports the applicability of MAAS across different fMRI input representations. However, \cite{park2025natural} identification score uses a different evaluation representation, and their models are subject-specific whereas MAAS uses a cross-subject approach. These values therefore provide contextual benchmarks rather than a controlled comparison of reconstruction methods. Overall, the experiment extends the evaluation of MAAS to an independent 3T dataset without establishing superiority or equivalence to the published pipeline.}

\paragraph{Auditory-Cortical Representation.}
\label{par:ac_repre}

The perceptual fidelity metric $\Delta_i$ is computed in the representational space of the NSL Auditory-Cortical Toolbox~\citep{chi2005multiresolution}, which models sound processing along the auditory pathway in two stages.

A waveform is decomposed into frequency-specific temporal envelopes using a bank of 128 asymmetric, constant-Q cochlear filters, spaced at 24 channels per octave relative to a 440~Hz reference frequency. Each channel then undergoes compression, lateral inhibition, and leaky temporal integration~\citep{chi2005multiresolution},
yielding the auditory spectrogram $y(t,x)$, a time--frequency representation reflecting auditory-nerve firing rates rather than raw acoustic energy.
The auditory spectrogram is then analysed by a bank of two-dimensional filters tuned to spectrotemporal modulations, modelling the selectivity of neurons in primary auditory cortex~\citep{kowalski1996analysis}. Each filter is
characterised by a temporal modulation rate $\omega$ and a spectral modulation scale $\Omega$:
\begin{equation}
    z(t,x;\omega,\Omega)
    =
    y(t,x) *_{t,x}
    \left[
        g_w(t;\omega)\, h_w(x;\Omega)
    \right].
\end{equation}

Upward and downward temporal-rate channels are averaged and integrated over time, yielding a three-dimensional modulation transfer function (MTF) of size $32~\text{scales} \times 64~\text{rates} \times 128~\text{frequency channels}$ per sound. For $\Delta_i$, Pearson correlation is computed between the flattened three-dimensional representations of reconstructed and ground-truth audio.

\begin{figure}
    \centering
    \includegraphics[width=.85\linewidth]{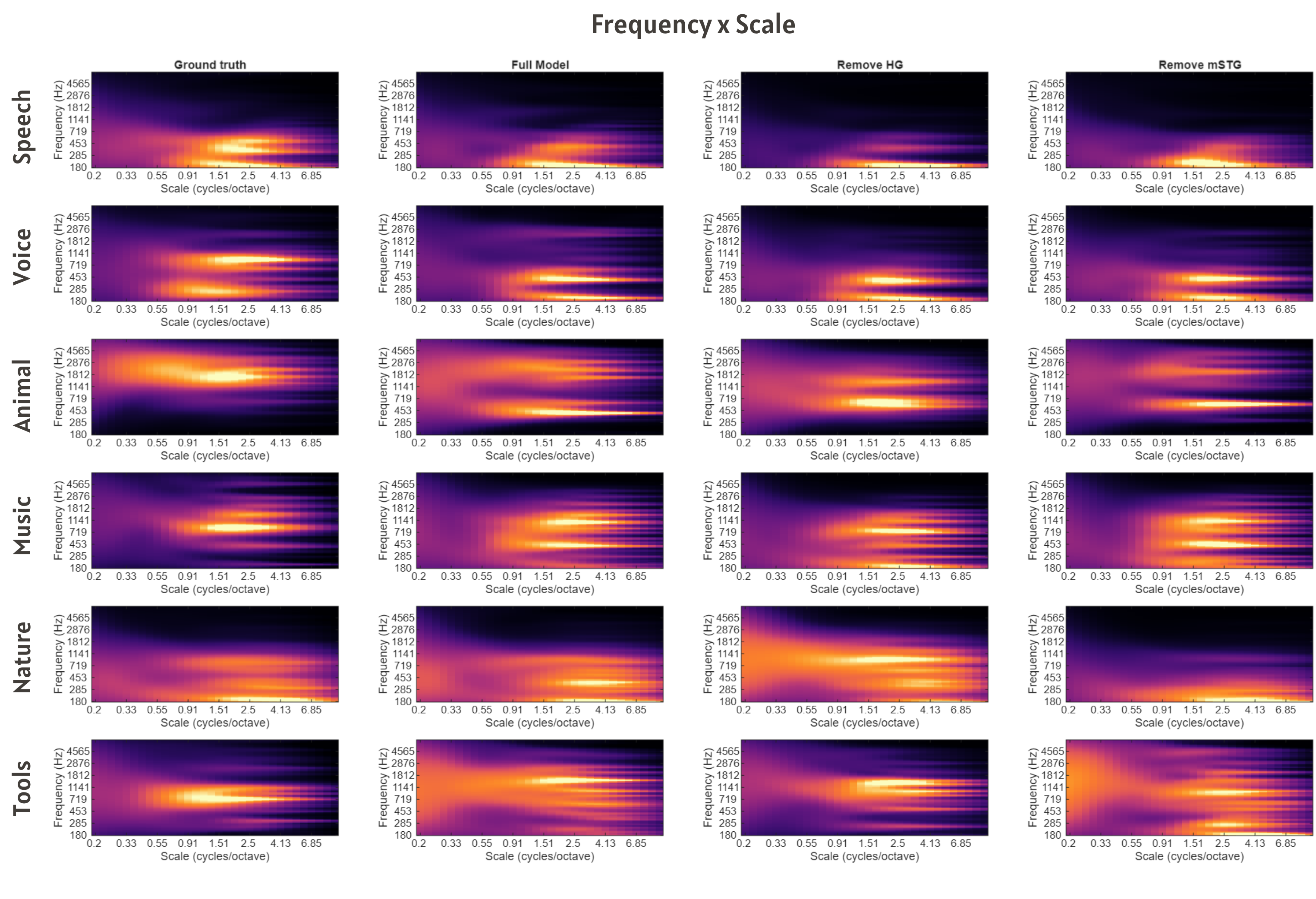}
    \caption{
    Category-averaged frequency--scale projections of NSL-derived
spectrotemporal representations, obtained by averaging over temporal
modulation rate. Rows correspond to semantic categories; columns show
the ground-truth audio, full-model reconstructions, and reconstructions
from models trained without HG or mSTG, respectively. The horizontal
axis denotes spectral modulation scale (cycles/octave), and the
vertical axis denotes characteristic frequency (Hz).
    }
    \label{fig:freqxscale}
\end{figure}

\begin{figure}
    \centering
    \includegraphics[width=.85\linewidth]{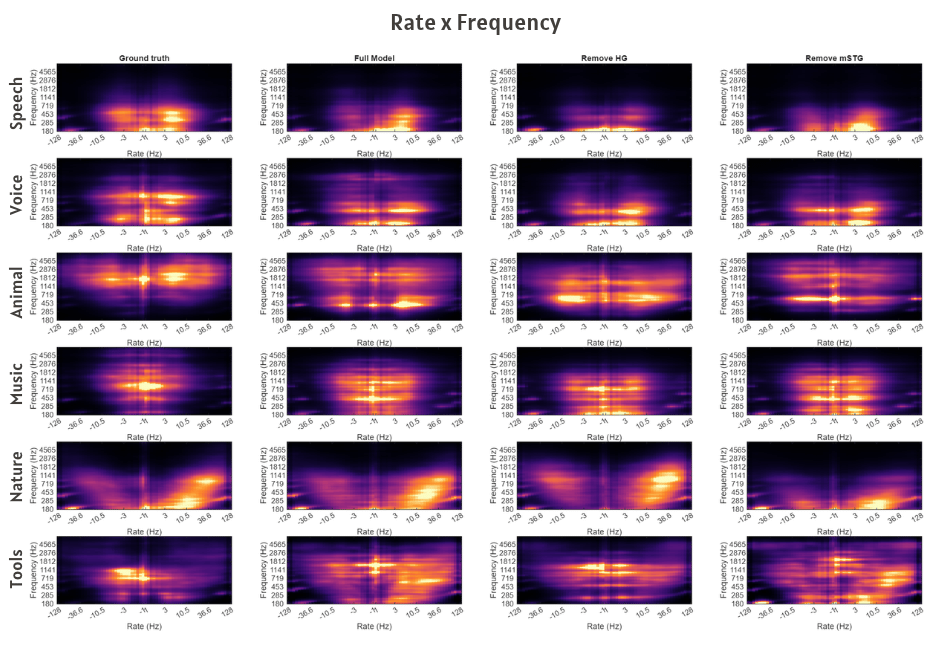}
    \caption{
    Category-averaged rate--frequency projections of NSL-derived
spectrotemporal representations, obtained by averaging over spectral
modulation scale. Rown and columns follow the same structure as frequency~$\times$~scale. The horizontal
axis denotes temporal modulation rate (Hz), and the vertical axis
denotes characteristic frequency (Hz).
    }
    \label{fig:ratexfreq}
\end{figure}

\begin{figure}
    \centering
    \includegraphics[width=.85\linewidth]{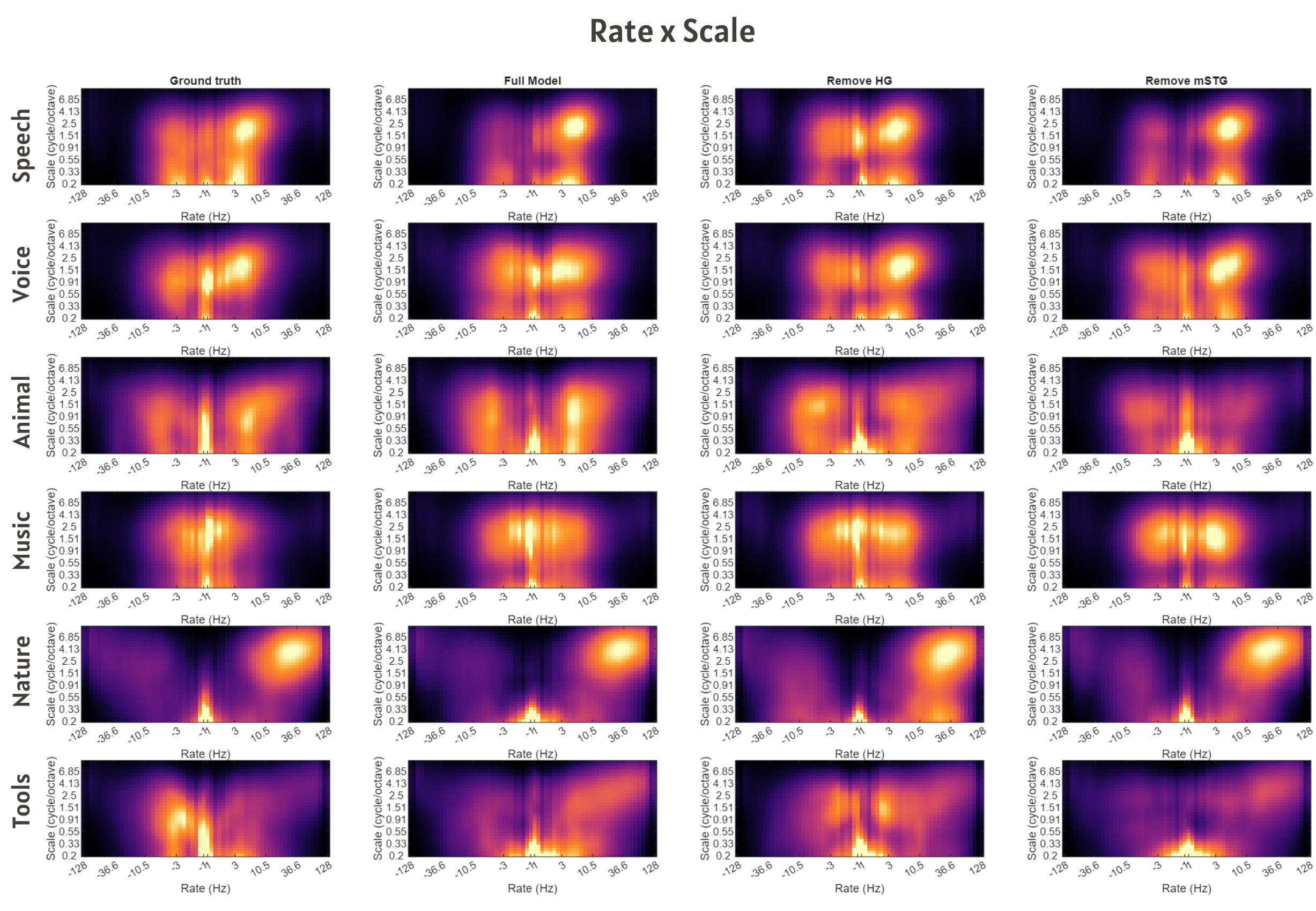}
    \caption{
    Category-averaged rate--scale projections of NSL-derived
spectrotemporal representations, obtained by averaging over
characteristic frequency. The horizontal axis denotes temporal modulation rate
(Hz), and the vertical axis denotes spectral modulation scale
(cycles/octave).
    }
    \label{fig:ratexscale}
\end{figure}

\textcolor{black}{For visualization, each three-dimensional MTF is projected into three two-dimensional summaries: rate~$\times$~scale (Figures~\ref{fig:ratexscale}), averaged over frequency; rate~$\times$~frequency (Figures~\ref{fig:ratexfreq}), averaged over scale; and frequency~$\times$~scale (Figures~\ref{fig:freqxscale}), averaged over rate. } These projections are computed for ground-truth sounds, the full model, and ablated or ROI-specific conditions, and visualised either for individual sounds or as category-averaged responses across the six semantic classes. Colour scaling is applied using the 1st--99th percentile range, either globally across panels or normalised per panel to the ground-truth range.

\end{document}